\documentclass[aps,prb,preprint,superscriptaddress]{revtex4}

\usepackage{graphicx}
\usepackage{dcolumn}
\usepackage{bm}

\begin{document}


\title{A Model for Dipolar Glass and Relaxor Ferroelectric Behavior}


\author{Ronald Fisch}
\affiliation{382 Willowbrook DR, North Brunswick, NJ 08902}


\date{\today}

\begin{abstract}
Heat bath Monte Carlo simulations have been used to study a 12-state discretized
Heisenberg model with a type of random field, for several values of the randomness
coupling parameter $h_R$.  The 12 states correspond to the [110] directions of a
cube.  Simple cubic lattices of size $128 \times 128 \times 128$ with periodic
boundary conditions were used, and 32 samples were studied for each value of $h_R$.
The model has the standard nonrandom two-spin exchange term with coupling energy $J$
and a field which adds an energy $h_R$ to two of the 12 spin states, chosen randomly
and independently at each site.  We provide results for the cases $h_R / J =$ -2.5,
-2.0, -1.5, 3.0 and 4.0.  For all these cases except $h_R / J =$ -2.5, we see an
apparently sharp phase transition at a temperature $T_c$ where the specific heat and
the longitudinal susceptibility are peaked.  At $T_c$, the behavior of the peak in the
structure factor, $S ({\bf k} )$, at small $|{\bf k}|$ is a straight line on a
log-log plot.  However, the value of the slope of this line is different for $h_R /J =$
-1.5 and 3.0 than it is for $h_R / J =$ -2.0 and 4.0.  We believe that the first two
cases are showing the behavior of a cubic fixed point in a weak random field, and the
behavior of the second two cases are showing the behavior of an isotropic fixed point
when the Imry-Ma length is smaller than the sample size.  Below $T_c$, these $L = 128$
samples show ferroelectric order, and this order rapidly becomes oriented along one of
the eight [111] directions as $T$ is reduced.  This rotation of the ordering direction
is caused by the cubic anisotropy.  For $h_R / J =$ -2.5, we do not see clear evidence
of a single well-defined $T_c$.

\end{abstract}


\maketitle


\section{INTRODUCTION}

The ${\bf O_{12}}$ model\cite{Fis98,Fis98b} is a model of discretized Heisenberg
spins, {\it i.e.} ${\bf O_{12}}$ is a discrete subgroup of ${\bf O(3)}$.  We will put
this model on a simple cubic lattice of size $L \times L \times L$, with periodic
boundary conditions.  It is now almost 25 years since the original Monte Carlo
study\cite{Fis98} of the three-dimensional (3D) random field $\bf O_{12}$ model.
The numerical results presented there, which used $L \le 64$, and studied four samples
at two nonzero values of $h_R$, are crude by current standards.  However, the reasons
given for why this model is worthy of study are just as valid now as they were then.
This model is designed to represent a ferroelectric dipolar glass,\cite{HKL90,BR92}
such as a random mixture of argon atoms and carbon monoxide molecules on a face-centered
cubic lattice in which the concentration of carbon monoxide is greater than the
percolation threshold.  It is also believed to be relevant\cite{HV76} for ferroelectric
phases in strongly disordered perovskite alloys, which are often referred to as relaxor ferroelectrics.\cite{SI54,BD83,WKG92,Kle93,PTB87,PB99,TGLR17}  From our point of view, the
essential element for relaxor ferroelectric behavior is the alloy disorder, rather than
a particular chemistry.  A cubic perovskite which has a supercell ordering of a few
perovskite unit cells in size should not be considered a relaxor ferroelectric,
regardless of its chemical composition.

In the current paper, we extend our recent results\cite{Fis22} on a particular
realization of this model.  We have used the same computer program as before,
although the procedures we follow have been slightly modified.  We continue to study
lattices of size $L = 128$, and we now present results for several values of the
random field coupling parameter $h_R$.  This gives a more complete picture of the
phase diagram for this model.

Our Hamiltonian starts with the classical Heisenberg exchange form of a ferromagnet:
\begin{equation}
  H_{ex} ~=~ - J \sum_{\langle ij \rangle} {\bf S}_{i} \cdot {\bf S}_{j}.
\end{equation}
Each spin ${\bf S}_{i}$ is a three-component dynamical variable which has twelve
allowed states, which are unit vectors pointing the [110] directions of a cube.
The $\langle ij \rangle$ indicates a sum over nearest neighbors on the simple
cubic lattice.  The symbol $H$, which denotes the uniform external field, does
not appear explicitly in our Hamiltonian, because we have set $H$ to zero.
There is no loss of generality in setting $J = 1$.  We will also set Boltzmann's
constant to 1.  This merely establishes the units of energy and temperature.
The restriction of the spin variables to ${\bf O_{12}}$ builds a
temperature-dependent cubic anisotropy into the model.  A useful review of the
effects of such a cubic anisotropy has been given by Pfeuty and Toulouse.\cite{PT77}
A recent paper by Aharony, Entin-Wohlman and Kudlis\cite{AEK22} gives a current
view of the problem.  It emphasizes the now well-established result\cite{Has22}
that, in the absence of the random field, there is a stable cubic critical point
in this model in 3D.  However, the isotropic ${\bf O(3)}$ is only slightly
unstable, and it has a strong influence on the finite-size scaling behavior.
The effect of a random field on this situation is what we will be trying to
understand in the work presented here.

We add to $H_{ex}$ a random field which, at each site, is the sum of two Potts
field terms:
\begin{equation}
  H_{RP2} ~=~ h_{R} \sum_{i} (\delta_{{\bf S}_{i} , {\bf \hat{u}}_{i}} +
\delta_{{\bf S}_{i} , {\bf \hat{v}}_{i}})   \, .
\end{equation}
Each ${\bf \hat{u}}_{i}$ and each ${\bf \hat{v}}_{i}$ is an independent quenched
random variable. which assumes one of the twelve [110] allowed directions with equal
probability.  Since ${\bf \hat{u}}_{i}$ is allowed to be equal to ${\bf \hat{v}}_{i}$,
the random field at each site has 78 possible types.  The reader should note that,
for an $XY$ model, adding a second Potts field would generate nothing new, due to the
Abelian nature of the ${\bf O(2)}$ group.  However, in the current case, adding the
second Potts field can substantially reduce the corrections to finite-size scaling.

Due to the way we have defined $H_{RP2}$, the energy per site of the system, $E$,
does not go to zero as the temperature, $T$, goes to infinity.  Actually, $E ( T =
\infty ) ~=~ h_R / 6$, and it would be straightforward to adjust the energy so that
it would go to zero as $T \to \infty$ for any value of $h_R$.  Although the reader
might think it would be more natural to include a minus sign in $H_{RP2}$, the
convention we are using is consistent with the definition used in our earlier work.\cite{Fis03}

The celebrated Imry-Ma construction,\cite{IM75} argues that random fields
built from groups with continuous symmetries cannot have conventional long-range
order when the number of spatial dimensions is less than or equal to four.
The Imry-Ma argument implicitly assumes that the spin group is Abelian.  What is
assumed is that the coupling of the random field to the spins is purely of vector
form. However, a nonabelian random vector field has the ability to generate higher
order tensor couplings to the effective spins under a rescaling transformation.
Fisher\cite{Fi85} has shown that for an isotropic distribution of random fields it
is necessary to keep track of all these higher order random tensor couplings
when the spatial dimension approaches four. Although our model is not isotropic,
the fact that the isotropic fixed point is close to stability for ${\bf O(3)}$ in 3D
means that we need to be concerned about what will happen to the relative stability of
the fixed points when the random field is added.  Even though we might expect that the
isotropic fixed point does not matter in the limit of a very weak random field, we
cannot assume that will continue to be true when the random field can no longer be
considered very weak.  When we consider nonlinear effects, it becomes unclear what
will happen even in the weak random field limit.  Since such nonlinear effects will
not be visible for small values of $L$, one should not expect that good finite-size
scaling will hold down to small values of $L$ in this situation.

Kleeman\cite{Kle93} has suggested that the random field Ising model (RFIM) is an adequate
approximation for the understanding of relaxor ferroelectric behavior.  It has recently
been shown\cite{KBPW22} that, on a simple cubic lattice at zero temperature, the critical
exponents of the random field 3-state Potts model are close those of the RFIM.  This supports
Kleeman's idea.  However, it can happen that a particular random field model has a specific
property which gives rise to special behavior.

In the work reported here, we show results for $h_R$ = -2.5, -2.0, -1.5, 3.0 and 4.0.
Note that having $h_R$ positive means that ${\bf S}_{i}$ has an increased energy if
it points in the direction of ${\bf \hat{u}}_{i}$ or ${\bf \hat{v}}_{i}$.  Thus for
negative $h_R$ there are usually 2 low energy states at each site, and for positive $h_R$
there are usually 10 low energy states at each site.  It follows that for $h_R$ values of
the same absolute value, the negative $h_R$ has a stronger effect on the system than the
positive $h_R$ does, because confining a spin to 2 directions out of 12 is a much
stronger restriction than confining a spin to 10 directions out of 12.

We expect, however, that the qualitative nature of the thermodynamic phases will not
depend on whether $h_R$ is positive or negative, or even whether the values of $h_R$
are equal for ${\bf \hat{u}}_{i}$ and ${\bf \hat{v}}_{i}$.  Our results seem to
support this expectation.  In any case, the size of corrections to scaling is not
universal.  The author expects that corrections to scaling will be smallest when
$h_R$ is the same for both ${\bf \hat{u}}_{i}$ and ${\bf \hat{v}}_{i}$.  Note that,
with this type of a random field, the model does not become trivial in the limit
$|h_R| \to \infty$.

The range of distributions of the random fields for which this effect will be found
remains unclear.  The pure Heisenberg critical point is not stable against the
introduction of cubic anisotropy which favors the [100] directions.  In that case,
the phase transition becomes first order.\cite{PT77}  There is no existing proof that
a first-order phase transition with a latent heat is impossible in a 3D model with a
quenched random field of the $H_{RP2}$ type.\cite{AW89,AW90,HB89}  However, an old
result of Blankschtein, Shapir and Aharony\cite{BSA84} argues that this will not occur
in Potts models.  Recent calculations\cite{KBPW22,TB21} on 3D Potts models with random
fields did not find any first-order phase transition behavior.

We will not find any evidence of [100] oriented magnetization in our model in the
average over samples.  However, unless the random field strength $h_R$ is weak, some
of the individual samples can show an average magnetization which is close to [100].
Since the random field disorder is not self-averaging, there is no reason why this
effect should disappear in the large $L$ limit.

\section{NUMERICAL PROCEDURES}

If $h_R$ is chosen to be an integer, then the energy of any state of the
model is an integer.  Then it becomes straightforward to design a heat-bath
Monte Carlo computer algorithm to study it which uses integer arithmetic
for the energies, and a look-up table to calculate Boltzmann factors.
This procedure substantially improves the performance of the computer
program, and was used for all the calculations reported here.  The program
currently has the ability to use half-integer values of $h_R$.  Lattices
with periodic boundary conditions were used throughout.

Three different linear congruential random number generators are used.
The first chooses the ${\bf \hat{u}}_{i}$, the second chooses the
${\bf \hat{v}}_{i}$ and the third is used in the Monte Carlo routine which
flips the spins.  The generator used for the spin flips, which needs very
long strings of random numbers, was the Intel library program $random\_number$.
In principle, Intel $random\_number$ can be used for multicore parallel
processing.  However, our program is so efficient in single-core mode that no
speedup was seen when the program was run in parallel mode.  The code was checked
by correctly reproducing the known results\cite{Fis98} of the $h_R = 0$ case,
and extending them to $L = 128$.

For each value of $h_R$, 32 samples of size $L = 128$ were studied.  The same set
of 32 samples was used for all values of $T$, so it is meaningful to talk about
heating and cooling of each sample.  Each sample was initially placed in a random
state, and then cooled slowly.  Both cooling and heating were done in temperature
steps of 0.015625, with times of 20,480 Monte Carlo steps per spin (MCS) at each
stage.  (The reader should note that the temperatures we use in the computer code
are simple binary fractions, even though they may appear to be unnatural when
written in decimal notation.)

For the pure ${\bf O_{12}}$ system, the Heisenberg critical temperature\cite{Fis98}
is $T_c$ = 1.453.  For $h_R = 3$, the spin correlations are only short-ranged
above this temperature, so no extended data runs were made in this region of $T$.
For 30 of the 32 samples studied, the system was strongly oriented along one
of the [111] directions by the time it had been cooled to $T$ = 1.3125.  The
other two samples had become trapped in metastable states.  These two samples
were then initialized in [110] states which had a positive overlap of approximately
0.5 with their metastable states.  They were run at $T$ = 1.25,  where they
were easily able to relax to low energy [111] states.  These states were
then warmed slowly to $T$ = 1.4375.  At $T$ = 1.34375, these [111] states were
seen to have significantly lower energies than the metastable states found by
cooling for those two samples.

The fact that it was not difficult to find a stable state oriented along [111]
at $T$ = 1.25 means that, for $h_R = 3$, an $L = 128$ sample does not have many
metastable states.  However, it is expected that for large enough values of $L$
it should be true that in the [111] ferroelectric phase there would be a
metastable state corresponding to each of the [111] directions.  It is to be
expected that this might no longer be true when $| h_R |$ becomes large.
In the proposed power-law correlated phase,\cite{Fis98} it would no longer be
possible to assign low-energy metastable states to particular [111] directions.

Trial runs made on $L = 64$ lattices allowed us to estimate the range of $T$
of most interest for each value of $h_R$ which was studied.  In the $h_R = 0$ case,
we know that there is a first order phase transition from a [111] ordered phase to
a [110] ordered phase.  However, a detailed study of the behavior of this second
transition was not undertaken.  At low temperatures, the presence of $h_R$ makes
it difficult to tell when the system has equilibrated.  In any case, the existence
of the lower transition depends on the details of the model, so it is probably
not relevant to experimental systems.

A data collection run for each sample consisted of a relaxation stage and a
data stage.  Except for $T$ values high enough so that correlations were much
smaller than the sample size, a relaxation stage was a run of length 122,880 MCS.
In many cases this was sufficient to bring the sample to an apparent local
minimum in the phase space.  This was followed by a data collection stage of the
same length.  If further relaxation was observed during the data collection stage,
it was reclassified as a relaxation stage.  Then a new data collection stage was
run.  The energy and magnetization of each sample were recorded every 20 MCS.
A spin state of the sample was saved after each 20,480 MCS.  Thus there were six
spin states saved from the data collection stage for each run.  These six spin
states were Fourier transformed and averaged to estimate the structure factor for
each sample.  At the high values of $T$, a relaxation stage of only 20,480 MCS was
judged to be sufficient.

\section{THERMODYNAMIC FUNCTIONS}

For a random field model, unlike a random bond model, the average value of
the local spin, $\langle {{\bf S}_i} \rangle$, is not zero even in the
high-temperature phase.  The angle brackets here denote a thermal average.
Thus the longitudinal part of the susceptibility,
$\chi_{||}$, is given by
\begin{equation}
  T \chi_{||} ({\bf k}) ~=~ 1 - |{\bf M}|^2 ~+~ L^{-3} \sum_{ i \ne j } \cos (
  {\bf k}  \cdot {\bf r}_{ij}) ({\bf S}_{i} \cdot {\bf S}_{j} ~-~ Q_{ij} )  \,   ,
\end{equation}
For Heisenberg spins,
\begin{equation}
  Q_{ij} ~=~ \langle {{\bf S}_i} \rangle \cdot \langle {{\bf S}_j} \rangle  \,  ,
\end{equation}
and
\begin{equation}
  |{\bf M}|^2 ~=~ L^{-3} \sum_{i} Q_{ii}
      ~=~ L^{-3} \sum_{i} [ \langle {\bf S}_{i} \rangle \cdot \langle {\bf S}_{i} \rangle ]_t \,  .
\end{equation}
where the square brackets $[ ... ]_t$ indicate a time average.  $Q_{ij}$ must be
included for all $T$.  Note that the $\chi_{||}$ we define here is not exactly what
would be called the longitudinal susceptibility in a nonrandom system.  The distinction
between longitudinal modes and transverse modes is not completely well-defined in a
system which has a local order parameter that is not fully aligned with the
sample-averaged order parameter.  In any case, below $T_c$, in the ferroelectric phase,
the cubic anisotropy suppresses the soft modes in our system.

When there is a phase transition, if the low temperature phase contains multiple Gibbs
states these equations are not adequate in the limit $L \to \infty$.  If multiple Gibbs
states exist at some $T$, then each $\langle {\bf S}_{i} \rangle$ must acquire an index
$\kappa$, which specifies the Gibbs state to which it belongs.  Each $Q_{ij}$ will
inherit this same index, {\it i.e.} $Q_{ij} ~\to~ Q^{\kappa}_{ij}$.  For our model,
$\kappa$ specifies one of the eight [111] directions.  Our data reported here are taken
at fixed $L = 128$, and we did not attempt to find systematically the local minima
corresponding to all the [111] directions.  This would not even be possible for $L = 128$
at $T$ only a small amount below the apparent $T_c$.  Under those conditions a sample can
often find its most stable local minimum within the time we run our Monte Carlo algorithm.
Thus we do not need to include $\kappa$ explicitly in the analysis of our data.

The structure factor, $S ({\bf k}) = \langle |{\bf M}({\bf k})|^2 \rangle $,
for Heisenberg spins is
\begin{equation}
  S ({\bf k}) ~=~  L^{-3} \sum_{ i,j } \cos ( {\bf k} \cdot {\bf r}_{ij})
   {\bf S}_{i} \cdot {\bf S}_{j}  \,   ,
\end{equation}
where ${\bf r}_{ij}$ is the vector on the lattice which starts at site $i$
and ends at site $j$.  When there is a phase transition into a state with
long-range spin order, $S ( {\bf k} = 0 )$ has a stronger divergence than
$\chi ( {\bf k} = 0 )$ does.  $S ({\bf k})$ is written using the basis of
eigenvectors of a translationally invariant system, even for a disordered
system, because this is the basis for which the results of X-ray scattering
and neutron scattering experimental results are measured.  Therefore, the
true eigenvalues of the $S ({\bf k})$ matrix in a disordered system are not
easily measured, especially when the disorder strength is large.  It is
necessary to remember this when interpreting the results reports here, as
well as when one compares to experimental measurements.

The specific heat, $c_H$, may be calculated by taking the finite differences
$\Delta E / \Delta T$, where $E$ is the energy per spin.  Alternatively, it
may be calculated as the variance of $E$ divided by $T^2$.  The second method
was used for the $c_H$ numbers shown in our tabulated data.

Define $M_x (t)$, $M_y (t)$ and $M_z (t)$ to be the averages over the lattice at
time $t$ of the components of ${\bf S}_i$, in the usual way.  Then the cubic
orientational order parameter (COO) is measured by calculating the quantity
\begin{equation}
  COO = 3 [ ({M_x}^2 {M_y}^2 + {M_x}^2 {M_z}^2 + {M_y}^2 {M_z}^2) / |{\bf M}|^4 ]_t   \, .
\end{equation}
where the brackets $[ ... ]_t$ indicate an average over time as well as an average
over samples.  The possible values of COO range from 0 when ${\bf M}$ points in a [100]
direction to 1 when ${\bf M}$ points along a [111] direction.  Note that if all of the
spins were fully aligned along one of the [110] directions, the value of COO would be
$3/4$.

\subsection{Weak Random Field Strength}

If the value of $| h_R |$ is very small, then it is necessary to use samples of very
large $L$, in order for the system to get through the crossover from pure system
behavior and show the true random field behavior.  It was found empirically that
$h_R = -1.5$ and $h_R = 3.0$ were satisfactory choices for $L = 128$ samples.  The
thermodynamic data calculated from our Monte Carlo data on the 32 $L = 128$ samples
for $h_R = 3$ are shown in Table I.  There are some small differences between these
data and the original tabulation shown for $h_R = 3$ in the earlier paper,\cite{Fis22}
because a stricter criterion for deciding that a sample had reached equilibrium was
used here.  This table now includes the results for both the hot start and cold start
runs at $T = 1.40625$.  In addition, for the case $T = 1.390625$, the prior report
accidently averaged over only 22 of the 32 samples which have been calculated.

The data in Table I show that there are peaks in $\chi_{||}$ and $c_H$ at $T_c$.
However, accurate estimates of the values of the correlation length critical exponent,
$\nu$, would require collecting a lot more data at temperatures close to $T_c$.  As
we shall see, our data below $T_c$ do not have the form of typical critical scaling
behavior.  This may be partly a reflection of the fact that this type of model is
expected to show replica-symmetry breaking\cite{MY92} RSB) below $T_c$.  It has been
understood for a long time that, in a mean-field approximation, RSB is a type of
ergodicity breaking.\cite{SZ82,Som83}  We are very far from mean-field theory here, however.

Table I shows that the sample average of COO is approximately 0.60 in the
paraelectric phase, and that it rapidly increases to 1 as $T$ is reduced below $T_c$.
This behavior indicates that above $T_c$ there is no significant preference for [111]
orientational ordering.

\begin{quote}
\begin{flushleft}
Table I: Thermodynamic data for $128 \times 128 \times 128$
lattices at $h_R = 3.0$, for various $T$. (h) and (c) signify
data obtained relaxing from hotter and colder initial conditions,
respectively.  The one $\sigma$ statistical errors shown are due
to the sample-to-sample variations.
\begin{tabular}{|l|ccccc|}
\hline
$~~~~~~T$&$|{\bf M}|$&$\chi_{||}$&$E$&$c_{H}$&COO\\
\hline
1.34375(c)&0.427$\pm$0.002&25.5$\pm$3.6&-1.1955$\pm$0.0001&2.784$\pm$0.009&0.985$\pm$0.006\\
1.375~~~(c)&0.311$\pm$0.004&86.0$\pm$7.8&-1.1047$\pm$0.0002&2.997$\pm$0.012&0.869$\pm$0.029\\
1.390625(c)&0.230$\pm$0.004&151$\pm$19&-1.0569$\pm$0.0001&3.111$\pm$0.016&0.722$\pm$0.041\\
1.40625(c)&0.125$\pm$0.005&304$\pm$37&-1.0072$\pm$0.0002&3.182$\pm$0.018&0.620$\pm$0.040\\
1.40625(h)&0.124$\pm$0.006&346$\pm$43&-1.0072$\pm$0.0002&3.184$\pm$0.023&0.611$\pm$0.040\\
1.4375~(h)&0.0238$\pm$0.0012&72.4$\pm$1.7&-0.9173$\pm$0.0001&2.472$\pm$0.009&0.599$\pm$0.012\\
\hline
\end{tabular}
\end{flushleft}
\end{quote}

Values of $S (|{\bf k}|)$ calculated by taking an angular average for each
sample and then averaging the results for the 32 $L = 128$ samples.  In Fig.~1,
the results for $S (|{\bf k}|)$ at the 15 smallest non-zero values of
$|{\bf k}|$ are shown as a function of $T$ using a log-log plot.  For larger
$|{\bf k}|$ the slope of $S$ becomes less negative, because the system is then
in the crossover region between the pure system fixed point and the random-field
fixed point. Note that the data shown for the cold initial condition and the hot
initial condition at $T$ = 1.40625 are virtually indistinguishable.  At lower
temperatures, two of the hot initial condition samples became trapped in
metastable states for times accessible to our calculations.  For this reason,
we do not show the hot initial condition data for $T < 1.40625$.

For the range of $|{\bf k}|$ shown in Fig.~1, the fact that the straight-line
fit to the data for $T$ = 1.40625 is essentially perfect is remarkable.  This
only a numerical accident, however, since we did not tune the temperature to
find this condition.  The fit of the hot initial condition data to a straight
line has a slope of $-2.788 \pm 0.014$, and the fit to the cold initial
condition data has a slope of $-2.786 \pm 0.016$.  Averaging these gives
\begin{equation}
  -(4 - \bar{\eta}) ~=~ -2.787 \pm 0.014  \,
\end{equation}
as the slope of $S(|{\bf k}|)$ on the log-log plot at the critical point
in the scaling region. We do not reduce our error estimate, because the
data for the different initial conditions on the same set of samples
cannot be considered to be statistically independent.  Thus, at $h_R =3$
\begin{equation}
  \bar{\eta} ~=~ 1.213 \pm 0.014  \ ,
\end{equation}
which is a reasonable value for this quantity.  The value of $\bar{\eta}$
should not be sensitive to varying $T$ by a small amount away from $T$ = 1.40625.
This is shown by the fact that the data in Fig.~1 for $T$ = 1.390625 run parallel
to the data for $T$ = 1.40625 over a range of $|{\bf k}|$, and for smaller
$|{\bf k}|$ the slope of $S(|{\bf k}|)$ becomes more negative.  What this means
is that $\bar{\eta}$ is not a continuous function of $T$.  We are not seeing a
critical phase of the Kosterlitz-Thouless type, for which $\bar{\eta}$ would be
decreasing continuously as $T$ decreases below $T_c$.

The thermodynamic data for $h_R = -1.5$, shown in Table II, are generally very
similar to the data for $h_R = 3$.  The value of $T_c$ now appears to be slightly
higher than $T$ = 1.40625.  We see again that there are peaks in $\chi_{||}$ and
$c_H$ at $T_c$, and that the COO parameter increases rapidly toward 1 as $T$
decreases below $T_c$.

\begin{quote}
\begin{flushleft}
Table II: Thermodynamic data for $128 \times 128 \times 128$
lattices at $h_R = -1.5$, for various $T$. (h) and (c) signify
data obtained relaxing from hotter and colder initial conditions,
respectively.  The one $\sigma$ statistical errors shown are due
to the sample-to-sample variations.
\begin{tabular}{|l|ccccc|}
\hline
$~~~~~~T$&$|{\bf M}|$&$\chi_{||}$&$E$&$c_{H}$&COO\\
\hline
1.359375(c)&0.407$\pm$0.002&24.6$\pm$2.1&-1.4483$\pm$0.0001&2.848$\pm$0.011&0.985$\pm$0.006\\
1.375~~~(c)&0.352$\pm$0.003&47.7$\pm$4.0&-1.4027$\pm$0.0001&2.983$\pm$0.010&0.947$\pm$0.011\\
1.390625(c)&0.277$\pm$0.004&92.2$\pm$7.7&-1.3550$\pm$0.0002&3.096$\pm$0.012&0.858$\pm$0.027\\
1.40625(c)&0.177$\pm$0.006&273$\pm$48&-1.3051$\pm$0.0002&3.233$\pm$0.019&0.721$\pm$0.034\\
1.421875(c)&0.073$\pm$0.004&240$\pm$9&-1.2554$\pm$0.0001&3.043$\pm$0.016&0.619$\pm$0.028\\
1.4375~(h)&0.0291$\pm$0.0015&95.1$\pm$3.3&-1.2117$\pm$0.0001&2.548$\pm$0.011&0.598$\pm$0.019\\
\hline
\end{tabular}
\end{flushleft}
\end{quote}

In Fig.~2, the results for $S (|{\bf k}|)$ at the 15 smallest non-zero values
of $|{\bf k}|$ for $h_R = -1.5$ are shown as a function of $T$ again using a
log-log plot.  The picture is similar to Fig.~1, except that the curvature in
$S (|{\bf k}|)$ for $T$ just below $T_c$ is less visible.  This, combined with
the slightly higher value of $T_c$, suggest that $L = 128$ may not be quite
large enough to complete the full crossover to random field critical behavior
when $h_R = -1.5$.

The fit of the hot initial condition data for $T = 1.40625$ to a straight
line has a slope of $-2.735 \pm 0.024$, and the fit to the cold initial
condition data has a slope of $-2.767 \pm 0.021$.  Averaging these gives
\begin{equation}
  -(4 - \bar{\eta}) ~=~ -2.751 \pm 0.021  \,
\end{equation}
as the slope of $S(|{\bf k}|)$ on the log-log plot at the critical point
in the scaling region.  Thus,
\begin{equation}
  \bar{\eta} ~=~ 1.249 \pm 0.021  \,
\end{equation}
at $h_R = -1.5$.  Although the agreement for the values of $\bar{\eta}$ for these two
values of $h_R$, -1.5 and 3.0, is pretty good, that is not convincing evidence of
$\bar{\eta}$ being independent of $h_R$, because these two values of $h_R$ were both
chosen to be about as small as we could expect to be able to get meaningful results
for at $L = 128$.  For a proper test of the universality of $\bar{\eta}$,
we need results for other values of $h_R$.

\subsection{Intermediate Random Field Strength}

Now we show results for $h_R = 4.0$ and $h_R = -2.0$, so that $| h_R |$ is
somewhat larger.  The thermodynamic data for $h_R = 4.0$, shown in Table III,
are qualitatively similar to the data for $h_R = 3.0$.  The value of $T_c$ now
appears to be somewhat lower than $T$ = 1.390625.  We see again that there are peaks
in $\chi_{||}$ and $c_H$ at $T_c$, and that the COO parameter increases toward 1 as
$T$ decreases below $T_c$.

\begin{quote}
\begin{flushleft}
Table III: Thermodynamic data for $128 \times 128 \times 128$
lattices at $h_R = 4.0$, for various $T$. (h) and (c) signify
data obtained relaxing from hotter and colder initial conditions,
respectively.  The one $\sigma$ statistical errors shown are due
to the sample-to-sample variations.
\begin{tabular}{|l|ccccc|}
\hline
$~~~~~~T$&$|{\bf M}|$&$\chi_{||}$&$E$&$c_{H}$&COO\\
\hline
1.34375(c)&0.345$\pm$0.005&59$\pm$5&-1.1821$\pm$0.0002&2.798$\pm$0.008&0.874$\pm$0.024\\
1.359375(c)&0.276$\pm$0.006&101$\pm$10&-1.1372$\pm$0.0002&2.880$\pm$0.014&0.763$\pm$0.040\\
1.375~~~(c)&0.203$\pm$0.006&146$\pm$10&-1.0912$\pm$0.0002&2.963$\pm$0.014&0.678$\pm$0.044\\
1.375~~~(h)&0.199$\pm$0.007&163$\pm$15&-1.0911$\pm$0.0002&2.958$\pm$0.013&0.660$\pm$0.045\\
1.390625(c)&0.115$\pm$0.006&183$\pm$17&-1.0437$\pm$0.0002&3.016$\pm$0.014&0.612$\pm$0.040\\
1.390625(h)&0.115$\pm$0.006&271$\pm$27&-1.0437$\pm$0.0002&3.030$\pm$0.014&0.599$\pm$0.039\\
1.40625(h)&0.057$\pm$0.004&160$\pm$6&-0.9969$\pm$0.0001&2.939$\pm$0.015&0.570$\pm$0.028\\
1.421875(h)&0.030$\pm$0.002&84$\pm$2&-0.9534$\pm$0.0001&2.628$\pm$0.011&0.598$\pm$0.017\\
\hline
\end{tabular}
\end{flushleft}
\end{quote}

In Fig.~3, the results for $S (|{\bf k}|)$ at the 15 smallest non-zero values
of $|{\bf k}|$ for $h_R = 4.0$ are shown as a function of $T$, again using a
log-log plot.  Our best estimate of $T_c$ for $h_R = 4.0$ is slightly below
$T = 1.390625$.  The slope of $S(|{\bf k}|)$ on the log-log plot for
$T = 0.390625$ is -2.848 $\pm$ 0.20 for the hot initial condition, and -2.850
$\pm$ 0.20 for the cold initial condition.  Averaging these results gives
\begin{equation}
  -(4 - \bar{\eta}) ~=~ -2.849 \pm 0.020  \,
\end{equation}
for the slope of the log-log plot and
\begin{equation}
  \bar{\eta} ~=~ 1.151 \pm 0.020  \,
\end{equation}
at $h_R = 4.0$.  This value is not consistent with the values of $\bar{\eta}$
which we found for the weak $h_R$ cases.

At $T = 1.375$ the Fig.~3 data for $S (|{\bf k}|)$ are close to a straight
line with a slope which is close to -3.0.  In Fig.~4, we show data for
$|{\bf k}|^3 S (|{\bf k}|)$ over a range of $|{\bf k}|$ on a linear plot,
for a range of $T$ around $T_c$.  It shows that below $T_c$ there a range
of $T$ for which $|{\bf k}|^3 S (|{\bf k}|)$ is approximately constant over
the range of $|{\bf k}|$ shown.  This range cannot extend all the way to
$|{\bf k}| \to 0$ because there is a sum rule for $S (|{\bf k}|)$.  In any
case, we expect that the crystalline cubic anisotropy must suppress $S (|{\bf k}|)$
for small enough $|{\bf k}|$.

Now we discuss the data for the case $h_R = -2.0$.  The thermodynamic data are
displayed in Table~IV.  For this case, we extend the range of data down to $T = 1.25$,
which is low enough so that the system's behavior is reaching the strong COO limit.
Our best estimate of $T_c$ for $h_R = -2.0$ is slightly below $T = 1.375$, where
the agreement between the data using the hot initial condition and the cold initial
condition is again fairly good.  We also show similar data for $T = 1.359375$, where
the agreement is not so good.  This is again caused by 2 of the 32 hot initial condition
samples becoming stuck in metastable states when the temperature is lowered below $T_c$.

\begin{quote}
\begin{flushleft}
Table IV: Thermodynamic data for $128 \times 128 \times 128$
lattices at $h_R = -2.0$, for various $T$. (h) and (c) signify
data obtained relaxing from hotter and colder initial conditions,
respectively.  The one $\sigma$ statistical errors shown are due
to the sample-to-sample variations.
\begin{tabular}{|l|ccccc|}
\hline
$~~~~~~T$&$|{\bf M}|$&$\chi_{||}$&$E$&$c_{H}$&COO\\
\hline
1.25~~~(c)&0.535$\pm$0.002&11.0$\pm$1.4&-1.7927$\pm$0.0001&2.282$\pm$0.007&0.996$\pm$0.001\\
1.28125(c)&0.472$\pm$0.003&23.2$\pm$2.0&-1.7189$\pm$0.0001&2.420$\pm$0.009&0.974$\pm$0.006\\
1.3125~(c)&0.390$\pm$0.004&52.4$\pm$6.1&-1.6401$\pm$0.0002&2.608$\pm$0.012&0.920$\pm$0.013\\
1.328125(c)&0.332$\pm$0.006&62.4$\pm$5.1&-1.5987$\pm$0.0002&2.676$\pm$0.011&0.816$\pm$0.037\\
1.34375(c)&0.269$\pm$0.007&99$\pm$10&-1.5559$\pm$0.0002&2.750$\pm$0.012&0.706$\pm$0.047\\
1.359375(c)&0.193$\pm$0.008&174$\pm$20&-1.5119$\pm$0.0002&2.842$\pm$0.013&0.691$\pm$0.039\\
1.359375(h)&0.187$\pm$0.009&162$\pm$17&-1.5118$\pm$0.0002&2.838$\pm$0.012&0.648$\pm$0.045\\
1.375~~~(c)&0.120$\pm$0.007&207$\pm$16&-1.4668$\pm$0.0001&2.901$\pm$0.013&0.642$\pm$0.036\\
1.375~~~(h)&0.120$\pm$0.007&215$\pm$17&-1.4668$\pm$0.0001&2.900$\pm$0.013&0.636$\pm$0.035\\
1.390625(h)&0.067$\pm$0.004&159$\pm$8&-1.4217$\pm$0.0001&2.840$\pm$0.011&0.611$\pm$0.032\\
1.40625(h)&0.037$\pm$0.002&92$\pm$2&-1.3785$\pm$0.0001&2.661$\pm$0.011&0.593$\pm$0.028\\
\hline
\end{tabular}
\end{flushleft}
\end{quote}

In Fig.~5, the results for $S (|{\bf k}|)$ at the 15 smallest non-zero values
of $|{\bf k}|$ for $h_R = -2.0$ are shown as a function of $T$, again using a
log-log plot.  Our best estimate of $T_c$ for $h_R = -2.0$ is slightly below
$T = 1.375$.  The slope of $S(|{\bf k}|)$ on the log-log plot for
$T = 0.390625$ is -2.869 $\pm$ 0.30 for the hot initial condition, and -2.862
$\pm$ 0.33 for the cold initial condition.  Averaging these results gives
\begin{equation}
  -(4 - \bar{\eta}) ~=~ -2.866 \pm 0.030  \,
\end{equation}
for the slope of the log-log plot and
\begin{equation}
  {\bar \eta} ~=~ 1.134 \pm 0.030  \,
\end{equation}
at $h_R = -2.0$.  Therefore, it appears to be well established that the value
of ${\bar\eta}$ is not independent of $h_R$ for this model.  However, the trend
for negative $h_R$ is the same as for positive $h_R$, although the rate of
change of ${\bar\eta}$ with $h_R$ is larger when $h_R$ is negative.  This
confirms our {\it a priori} expectations.

The author cannot resist the observation that this effect reminds him of the
Angell plot,\cite{Ang95} which shows that the freezing behavior of ``strong
glasses" differs from the behavior of ``fragile glasses".  (Note that the Angell
plot uses the symbol $\eta$ for viscosity, which is not related to our notation.)

In Fig.~6, we show the $h_R = -2.0$ data for $|{\bf k}|^3 S (|{\bf k}|)$ over a
range of $|{\bf k}|$ on a linear plot.  It shows that below $T_c$ there a range
of $T$ for which $|{\bf k}|^3 S (|{\bf k}|)$ is approximately constant over
a range of $|{\bf k}|$, as in Fig.~4.  For lower $T$ the slope is positive,
because the long-range order is being stabilized by the COO.  We expect that
Fig.~4 would also show this stabilization if we had data for lower values of $T$
in the $h_R = 4.0$ case.

\subsection{Strong Random Field Strength}

For large positive $h_R$, $H_{RP2}$ acts like a constraint which forbids 2 of the 12
states, chosen randomly at each site, from being occupied.  Since 2 is much smaller
than 12, this constraint is probably not strong enough to destroy long-range order at
low temperatures. The situation for large negative $h_R$ is very different.  In that
case, only 2 of the 12 states on each site are allowed.  Such a model does not have
the Kramers degeneracy, and it is not expected to be in a spin-glass universality class.
However, this does not mean it would be trivial to find the ground state(s) of the model.
It is known that a necessary condition for the existence of a finite temperature phase
transition into a state which possesses true long-range order is that there be more than
one thermodynamic Gibbs state in the low temperature phase.  In the absence of Kramers
degeneracy, it is not clear that this condition can be satisfied.  On the other hand,
there is no proper proof that a glass-type phase transition cannot exist in our model at
finite $T$ in 3D. This issue is very similar to the question of whether the de
Almeida-Thouless line exists in a 3D Ising spin-glass,\cite{DAT78} or the Gabay-Toulouse
line exists in a Heisenberg spin-glass.\cite{GT81}

Thermodynamic data for $h_R = -2.5$ are displayed in Table~V.  We do not show data for
$T < 1.25$, because we do not have confidence in the ability of our calculations to
provide meaningful results within accessible relaxation times at lower $T$.  What these
data show is that there is no single value of $T$ which can be identified as $T_c$.  The
peak in $\chi_{||}$ is not occurring at the same $T$ as the peak in $c_{H}$.  The COO does
not start increasing until a even lower $T$ near 1.3 is reached.  We would need a larger
value of $L$ to be able to know if the COO goes to 1 for $h_R = -2.5$.  The author is not
able to do such a calculation at this time.

\begin{quote}
\begin{flushleft}
Table V: Thermodynamic data for $128 \times 128 \times 128$
lattices at $h_R = -2.5$, for various $T$. (h) and (c) signify
data obtained relaxing from hotter and colder initial conditions,
respectively.  The one $\sigma$ statistical errors shown are due
to the sample-to-sample variations.
\begin{tabular}{|l|ccccc|}
\hline
$~~~~~~T$&$|{\bf M}|$&$\chi_{||}$&$E$&$c_{H}$&COO\\
\hline
1.25~~~(c)&0.311$\pm$0.007&56$\pm$5&-1.8686$\pm$0.0002&2.295$\pm$0.008&0.721$\pm$0.033\\
1.28125(c)&0.204$\pm$0.011&84$\pm$6&-1.7941$\pm$0.0002&2.419$\pm$0.009&0.672$\pm$0.039\\
1.296875(c)&0.159$\pm$0.010&110$\pm$8&-1.7555$\pm$0.0002&2.483$\pm$0.010&0.648$\pm$0.040\\
1.3125~(c)&0.121$\pm$0.007&125$\pm$7&-1.7160$\pm$0.0001&2.535$\pm$0.008&0.580$\pm$0.044\\
1.328125(c)&0.085$\pm$0.006&134$\pm$8&-1.6757$\pm$0.0001&2.608$\pm$0.011&0.588$\pm$0.039\\
1.34375(h)&0.059$\pm$0.004&109$\pm$5&-1.6350$\pm$0.0001&2.620$\pm$0.009&0.580$\pm$0.035\\
1.359375(h)&0.041$\pm$0.003&76$\pm$2&-1.5945$\pm$0.0001&2.562$\pm$0.011&0.583$\pm$0.034\\
\hline
\end{tabular}
\end{flushleft}
\end{quote}

We did not find any value of $T$ for which the $h_R = -2.5$ data for $S (|{\bf k}|)$ looked like
a good straight line on a log-log plot, which is consistent with the lack of a well-defined $T_c$
in Table~V.  It is clear that the entire sample is strongly correlated at $T = 1.25$, but this
does not prove true long-range order for $L \to \infty$.  In Fig.~7 and Fig.~8, we show how
$|{\bf k}|^3 S (|{\bf k}|)$ versus $|{\bf k}|$ changes with $T$.  The hot initial condition
data of Fig.~7 show that $|{\bf k}|^3 S (|{\bf k}|)$ increases as $T$ decreases, but the slope
of the data at small $|{\bf k}|$ remains positive.  The data for the cold initial condition,
which extend down to $T = 1.25$, show a more complicated picture.  The small $|{\bf k}|$ data
for $T = 1.25$ are no longer rising as $T$ decreases.  This most likely indicates merely that
we would need a larger value of $L$ to see the true small $|{\bf k}|$ behavior at this $T$.

\section{DISCUSSION}

Our results for the cases with weak random fields, $h_R = -1.5$ and $h_R = 3.0$, can be
understood as showing ferromagnetism induced by the presence of the crystalline anisotropy.
This behavior is not surprising.  However, merely adding a weak random field will not
explain the behavior of relaxor ferroelectrics.  The cases with stronger random fields
require a deeper understanding.  In the absence of the random field, the stability
advantage of the cubic fixed point over the isotropic fixed point is very small.\cite{Has22}
This unusual situation means that adding a random field which is not very weak can
cause complex effects which would normally only be expected near a multicritical point.

What we would expect is that when the random fields become strong enough, the cubic
crystalline anisotropy will no longer dominate the observed behavior.  It is conceivable
that the two competing fixed points become entangled in a way which results in behavior
of a type which is completely new.  However, the numerical results presented here are
not sufficiently detailed to show that something that exotic is happening.

The simplest possibility for what happens for random fields which are not weak is that there
is a multicritical point as a function of the strength of the random field.  Beyond the
multicrical point, we need to introduce the concept of the Imry-Ma length scale.\cite{IM75}
The cases with random fields of intermediate strengths, $h_R = -2.0$ and $h_R = 4.0$, may be
understood as a situation where the random fields are able to dominate the over the crystalline
anisotropy, but the Imry-Ma length is larger than the size of the finite samples which we have
been able to study.  The typical sample then behaves like a single Imry-Ma domain.

For the case $h_R = -2.5$, the Imry-Ma length is now smaller than the size of the samples
which we have studied.  If we were able to study much larger samples, then we would expect
the intermediate random field cases would show the same qualitative behavior that we see
for $h_R = -2.5$.  Naturally, since we have not actually done that calculation, this is only
an extrapolation.  However, the behavior we see for $h_R = -2.5$ is not merely the exponential
decay of spin correlations which one expects to find in the limit that the random field is so
strong that it almost completely determines the direction in which each spin is forced to point.
It may be true that all spin correlations decay exponentially at any $T$ for large enough sample
sizes, but that is not what we are seeing.  There is still something nontrivial going on here
as $T$ is decreased, which our data are not adequate to show convincingly.

The naive interpretation of the Imry-Ma argument in 3D for ${\bf O (3)}$ spins without any
crystalline anisotropy and with an isotropic distribution of random fields is that the domain
structure of the low $T$ state should be uniquely determined for each sample by its own particular
set of random fields when $L$ is much larger than the Imry-Ma length, even when $| h_R / J | \ll 1$.
This, however, is not known to be correct.  It is not what is proven by Aizenman and Wehr.\cite{AW90}
What Aizenman and Wehr argue is that it is necessary for the existence of a ferromagnetic phase under
these conditions that in the limit $L \to \infty$ there exists a manifold of ferromagnetic Gibbs
states which is invariant under ${\bf O (3)}$ rotations.  This condition would be necessary to insure
that samples be self-averaging in the limit $L \to \infty$.  However, the reason why self-averaging
should hold for this problem is far from clear.  Mezard and Young\cite{MY92} argued in 1992 that it is
not true in less than six space dimensions.  The current author does not know of any proof that there
should be such a rotation-invariant manifold of ferromagnetic Gibbs states in 5D, where everyone seems to
agree that a ferromagnetic phase exists for weak $h_R / J$.  The essential point is that the
$\langle {{\bf S}_i} \rangle$ for a ferromagnetic state in a sample with a random field are not collinear,
even if the probability distribution for the random field is isotropic.  This means that the magnetic
susceptibility cannot be simply separated into a transverse part and a longitudinal part.  The
$``$transverse" magnetic susceptibility is thus not required to be infinite below $T_c$, contrary to what
is claimed by Aizenman and Wehr.\cite{AW90}

The procedures we follow may be thought of as a hybrid of ${\bf \vec{k}}$-space renormalization group
ideas and finite-size scaling.  The connection is made through the use of the fast Fourier transform,
which requires that $L$ be a power of two.  Looking at the small-$|{\bf k}|$ part of $S (|{\bf k}|)$
allows us to select out the part of the Monte Carlo data which contains information about the
long-wavelength behavior. The crossover from pure system behavior to random-field behavior, which dominates
the shorter wavelengths, does not need to have a simple correction-to-scaling functional form which can be
modeled by a critical exponent.  {\it A priori}, we are not assuming anything about whether we will find
some kind of a critical point.  As we have seen, no clear critical behavior was found in our calculation for
the case $h_R = -2.5$.  It is no surprise that ${\bf \vec{k}}$-space ideas become less useful when the
strength of the random field grows.

In order to understand the significance of our results, it is necessary to explain carefully how the
RP2 ${\bf O}_{12}$ model is related to the random-field Ising model (RFIM).  It was shown by Aharony\cite{Aha78}
that a small uniform uniaxial will cause an isotropic or cubic model in a random field to cross over to
the RFIM universality class.  Extensive numerical calculati ons by Fytas and Martin-Mayor\cite{FMM13,FMM16}
have shown that at $T_c = 0$ the 3D RFIM has a well-behaved critical point which obeys universality.  It must
be pointed out, however, that what Fytas and Martin-Mayor did was show that the fourth cumulant of the random
field probability distribution was an irrelevant variable at the RFIM critical point, and then calculate the
critical exponents for that case.  Generalizing the ideas of Lubensky,\cite{Lub75} we expect that all
cumulants of the fixed point random-field probability distribution need to be considered.

For example, the author does not believe it is obvious that a Cauchy distribution should give the same critical
exponents for the RFIM as the ones found by Fytas and Martin-Mayor.  Even more interesting is the situation for
a random field distribution which is not symmetric, so that it breaks the $Z_2$ symmetry.  A first-order phase
transition for the RFIM in 3D is allowed,\cite{AW89,AW90,HB89} in principle.  However, to the author's knowledge
there is no evidence of one, either numerical or experimental.  One should not expect universality to hold for
the 3D RFIM if the random-field distribution breaks the $Z_2$ symmetry.  This seems only slightly more unusual
than a chiral crystal, and less exotic than a quasicrystal.

In addition to those considerations, there is also another mechanism for the apparent breakdown of universality
which is special to the case of 3D ${\bf O(3)}$ models with cubic symmetry.  It is that the cubic fixed point
is only very slightly more stable than the isotropic fixed point.  If we apply a weak random field, it should
be expected that the flows in Hamiltonian space under rescaling transformations will be complicated.  The two
fixed points should interfere with each other, and the nature of this interference ought to depend on the
strength of the random field.  It is even possible that, when the random field becomes strong enough, evanescent
scaling\cite{PT77b} will occur.  This may be the explanation of our results for the $h_R$ = -2.5 case.

Thus it should not be considered a surprise that our results for $\bar{\eta}$ for the RP2 ${\bf O}_{12}$
model do not appear to obey universality.  In this context, it should be noted that Hui and Berker\cite{HB89}
claimed that the phase transition in any 3D random model should be a critical point, since such a model should
not have a first-order transition.  However, it is not clear that Hui and Berker intended to imply that such
a phase transition should have universal critical exponents.  Turkoglu and Berker\cite{TB21} have revisited
the random field Potts models, but they do not calculate critical exponents.  A recent calculation of critical
exponents for the model has been given by Kumar, Banerjee, Puri and Weigel,\cite{KBPW22} who use the
zero-temperature method.  The latter calculation studies only one type of random field, so there is no test
for universality of these critical exponents.

For the cases of weak and moderate $h_R$, where we see a well-defined $T_c$ and an ordering below $T_c$ which
lies close to one of the [111] axes, the behavior of our model is essentially that of a 3D $N = 3$ Ashkin-Teller
model in a random field.  However, this is different from the $N = 3$ Ashkin-Teller model with bond disorder.
It is the random bond version which has been related\cite{Car96,Car99} to the RFIM.  Much less is known about
the random field $N = 3$ Ashkin-Teller model, and there is no reason the author is aware of to expect that it
is as closely related to the RFIM.

\section{CONCLUSION}

In this work we have used Monte Carlo computer simulations to study a model
of discretized Heisenberg spins on simple cubic lattices in 3D which has a weak
cubic crystalline anisotropy.  It has a carefully chosen type of random field
which has a probability distribution with cubic symmetry.  We have found that,
if the strength of the random field is small, our model shows a well-defined
$T_c$ and a transition into a phase with long-range order oriented along one of
the [111] cubic axes.  This phase transition shows characteristics expected of
random field behavior.  However, the value of the apparent critical exponent
$\bar{\eta}$, which describes the scaling of $S (|{\bf k}|)$ at $T_c$ changes
when $|h_R|$ becomes somewhat larger.  We believe this breakdown of typical
scaling behavior is caused by the destabilization of the cubic fixed point by
the random field.  The behavior of the samples then appears to be that of an
isotropic ${\bf O(3)}$ random field model for which the Imry-Ma length is larger
than our sample size.

For the strongest random field case that has been studied here, a single
well-defined $T_c$ was not found.  Additional study is required to determine
if the proper interpretation of this is that there are now two distinct
phase transitions, or if the phase transition seen for weaker random fields
has become smeared out.  We have also identified the relaxor ferroelectric
effect as arising from the random field becoming strong enough to destabilize
the random field cubic critical fixed point, and the noncollinearity of the local
ferroelectric order parameter near $T_c$.

\begin{acknowledgments}

This work used the Extreme Science and Engineering Discovery Environment (XSEDE)
through allocation DMR180003.  Bridges Regular Memory at the Pittsburgh
Supercomputer Center was used for some preliminary code development, and the
new Bridges-2 Regular Memory machine was used to obtain the data discussed here.
The author thanks the staff of the PSC for their help.  The author's ideas about
the behavior of random field models have been clarified by recent discussions with
Amnon Aharony and Ron Peled.

\end{acknowledgments}


\newpage
\begin{figure}
\includegraphics[width=3.4in]{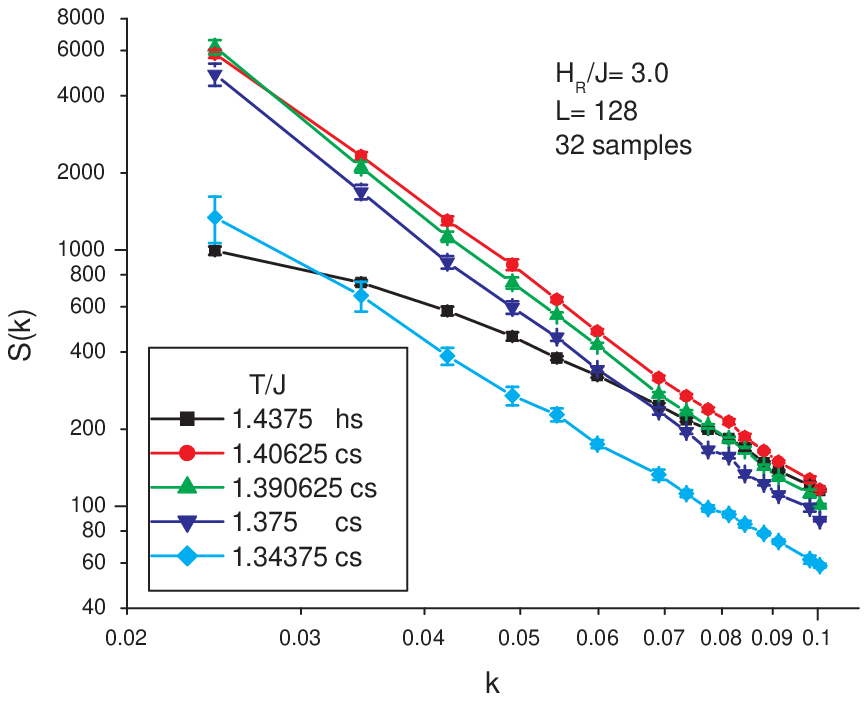}
\caption{\label{fig1} Angle-averaged structure factor, $S (|{\bf k}|)$,
at a sequence of temperatures for the RP2 ${\bf O}_{12}$ model with $h_R = 3.0$
on $128 \times 128 \times 128$ simple cubic lattices, log-log plot.  The points
show averaged data from 32 samples, at a series of temperatures. (hs) and (cs)
signify data obtained by relaxing from hotter and colder initial conditions,
respectively.}
\end{figure}

\begin{figure}
\includegraphics[width=3.4in]{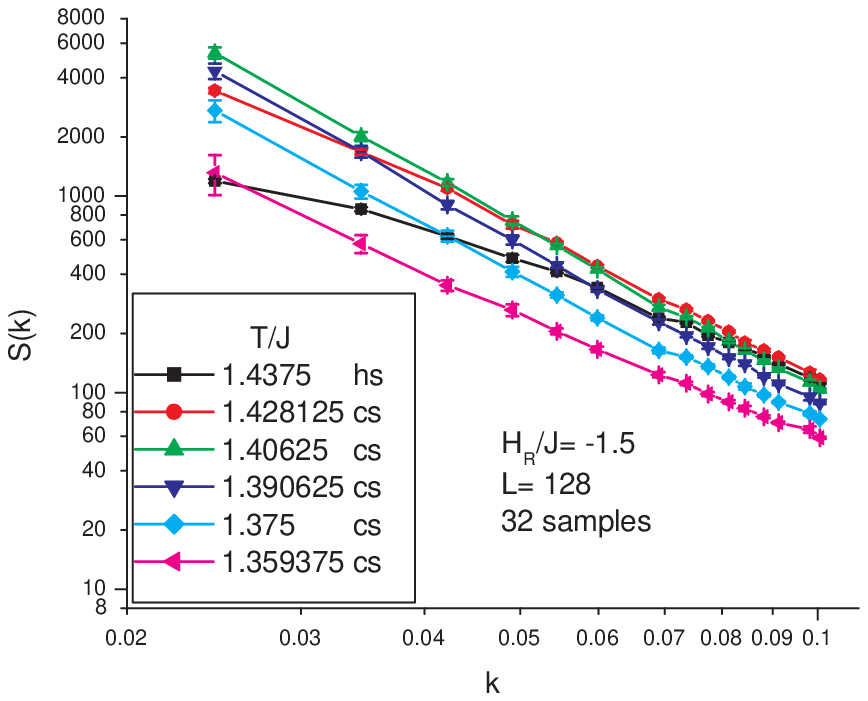}
\caption{\label{fig2} Angle-averaged structure factor, $S (|{\bf k}|)$,
at a sequence of temperatures for the RP2 ${\bf O}_{12}$ model
with $h_R = -1.5$ on $128 \times 128 \times 128$ simple cubic lattices, log-log plot.
The points show averaged data from 32 samples.  (hs) and (cs) signify data obtained
by relaxing from hotter and colder initial conditions, respectively.}
\end{figure}

\begin{figure}
\includegraphics[width=3.4in]{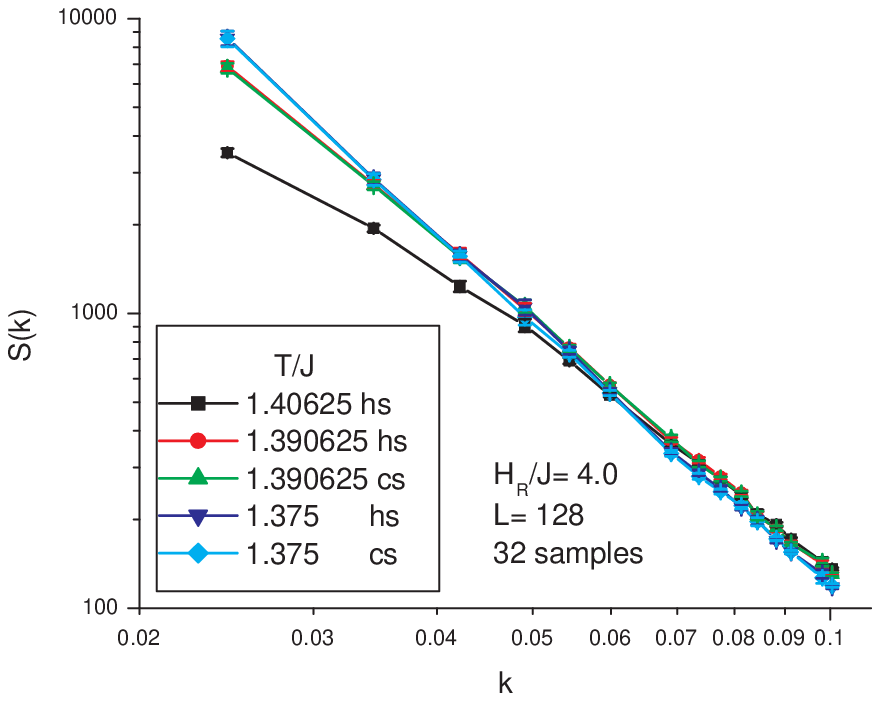}
\caption{\label{fig3} Angle-averaged structure factor, $S (|{\bf k}|)$,
at a sequence of temperatures for the RP2 ${\bf O}_{12}$ model with $h_R = 4.0$
on $128 \times 128 \times 128$ simple cubic lattices, log-log plot.  The points
show averaged data from 32 samples, at a series of temperatures. (hs) and (cs)
signify data obtained by relaxing from hotter and colder initial conditions,
respectively.}
\end{figure}

\begin{figure}
\includegraphics[width=3.4in]{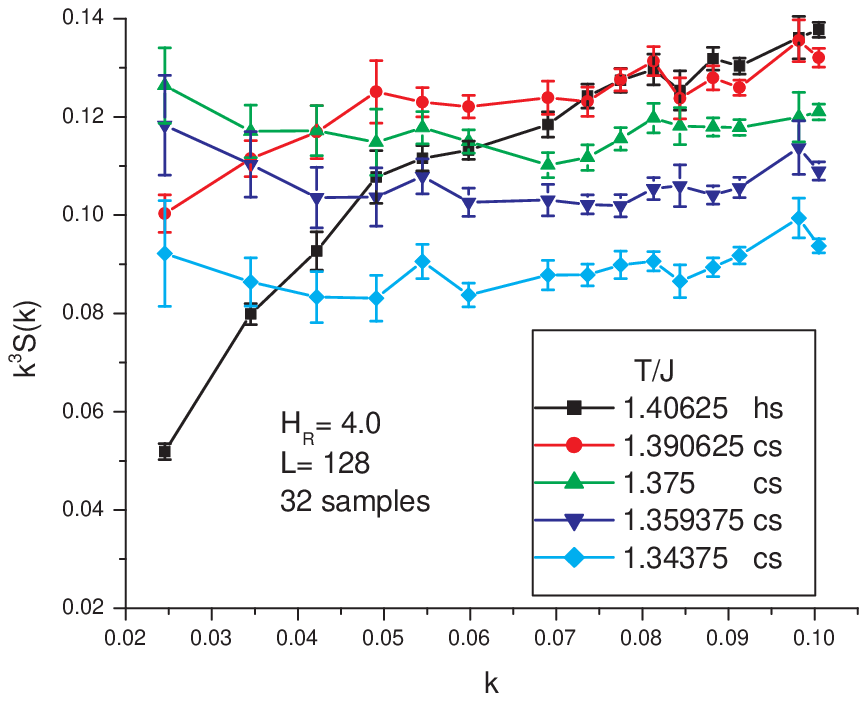}
\caption{\label{fig4} Angle-averaged structure factor, $S (|{\bf k}|)$,
at a sequence of temperatures for the RP2 ${\bf O}_{12}$ model
with $h_R = 4.0$ on $128 \times 128 \times 128$ simple cubic lattices.
The points show averaged data from 32 samples.  (hs) and (cs) signify data obtained
by relaxing from hotter and colder initial conditions, respectively.}
\end{figure}

\begin{figure}
\includegraphics[width=3.4in]{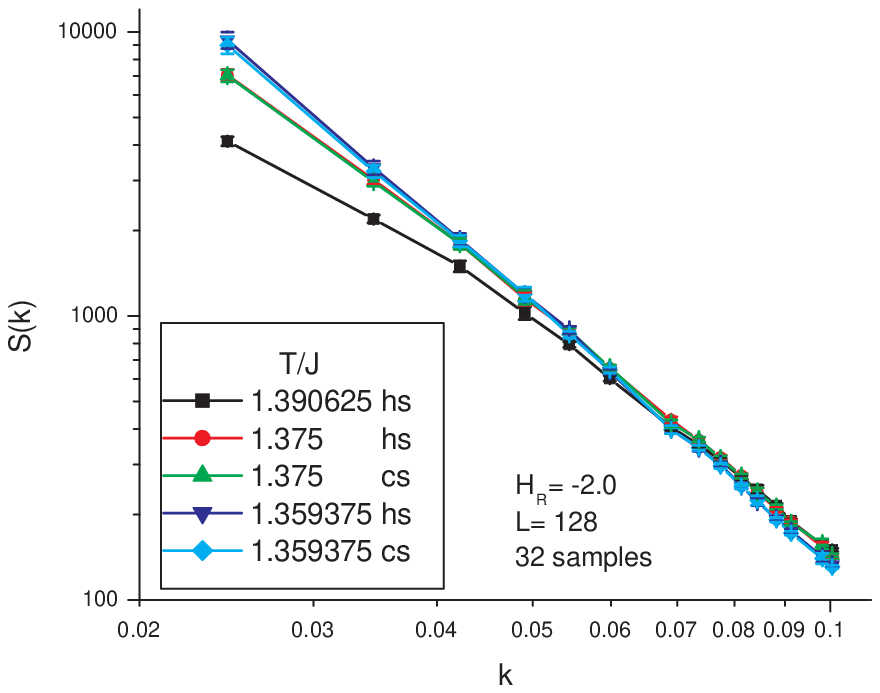}
\caption{\label{fig5} Angle-averaged structure factor, $S (|{\bf k}|)$,
at a sequence of temperatures for the RP2 ${\bf O}_{12}$ model with $h_R = -2.0$
on $128 \times 128 \times 128$ simple cubic lattices, log-log plot.  The points
show averaged data from 32 samples, at a series of temperatures. (hs) and (cs)
signify data obtained by relaxing from hotter and colder initial conditions,
respectively.}
\end{figure}

\begin{figure}
\includegraphics[width=3.4in]{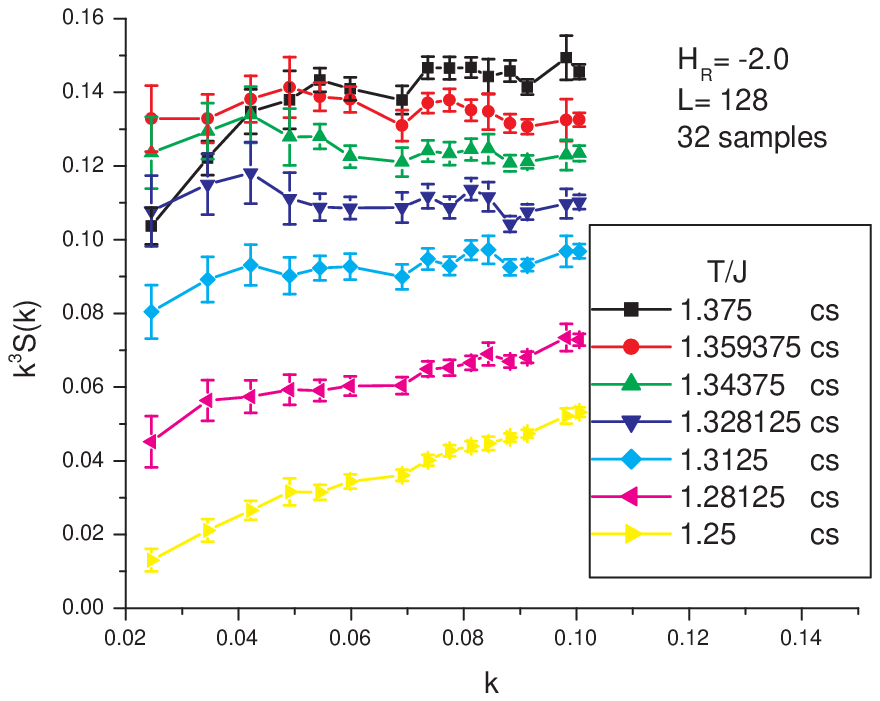}
\caption{\label{fig6} Angle-averaged structure factor, $S (|{\bf k}|)$,
at a sequence of temperatures for the RP2 ${\bf O}_{12}$ model
with $h_R = -2.0$ on $128 \times 128 \times 128$ simple cubic lattices.
The points show averaged data from 32 samples.  (hs) and (cs) signify data obtained
by relaxing from hotter and colder initial conditions, respectively.}
\end{figure}

\begin{figure}
\includegraphics[width=3.4in]{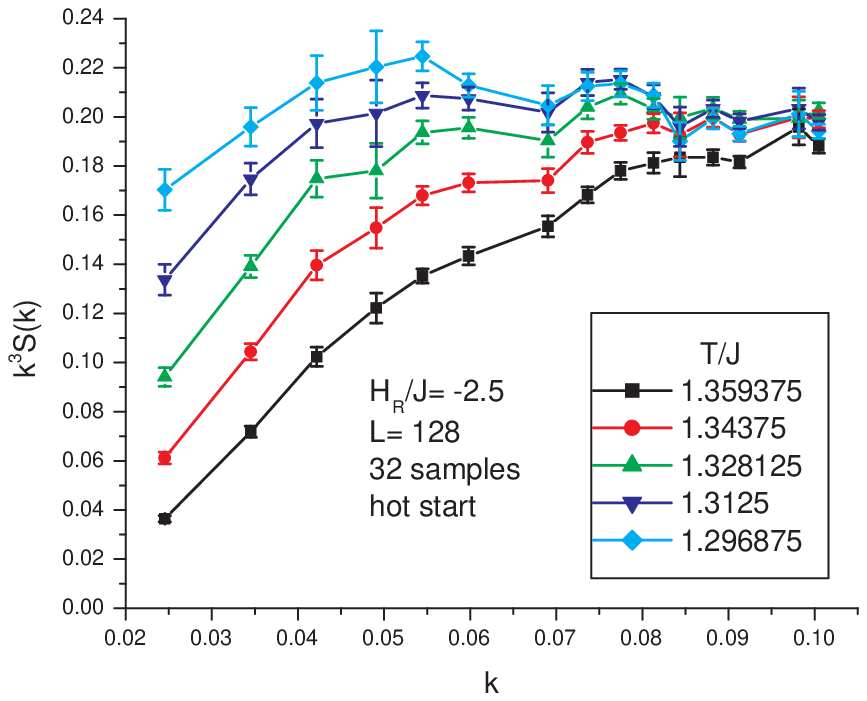}
\caption{\label{fig7} Angle-averaged structure factor, $S (|{\bf k}|)$,
at a sequence of temperatures for the RP2 ${\bf O}_{12}$ model
with $h_R = -2.5$ on $128 \times 128 \times 128$ simple cubic lattices.
The points show averaged data from 32 samples.  These data were obtained
by relaxing from hotter initial conditions.}
\end{figure}

\begin{figure}
\includegraphics[width=3.4in]{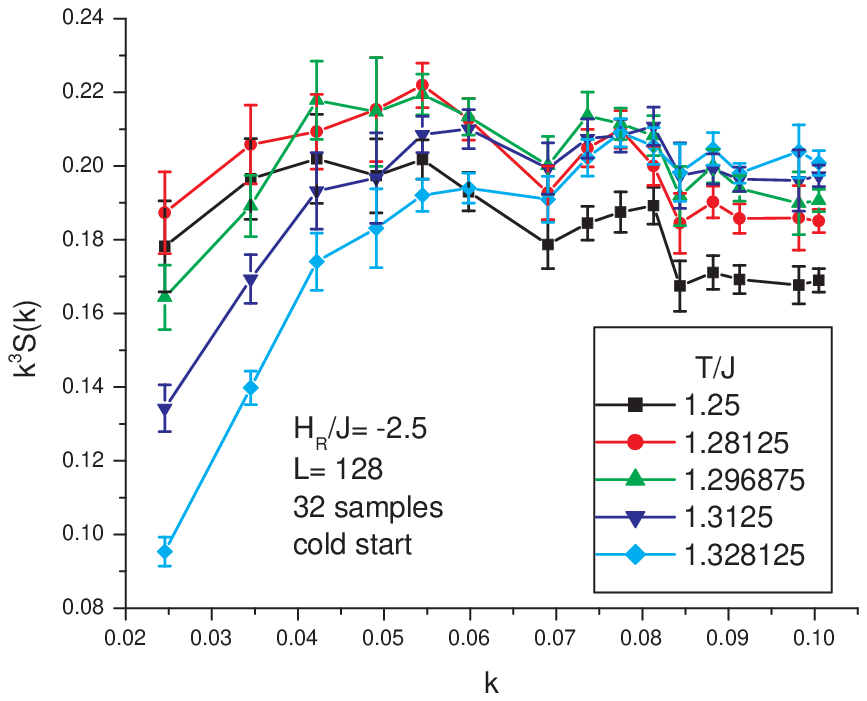}
\caption{\label{fig8} Angle-averaged structure factor, $S (|{\bf k}|)$,
at a sequence of temperatures for the RP2 ${\bf O}_{12}$ model
with $h_R = -2.5$ on $128 \times 128 \times 128$ simple cubic lattices.
The points show averaged data from 32 samples.  These data were obtained
by relaxing from colder initial conditions.}
\end{figure}

\end{document}